\documentclass[9pt,twocolumn,twoside,lineno]{pnas-new}
\usepackage[version=3]{mhchem} %for chem formula

\newcommand{\rev}[2]{#2}

\templatetype{pnasresearcharticle} % Choose template 
\title{3D tomographic analysis of the order-disorder interplay in the \textit{Pachyrhynchus congestus mirabilis} weevil}
\author[a]{Kenza Djeghdi}
\author[a]{Ullrich Steiner}
\author[a,b,1]{Bodo D. Wilts} 
\affil[a]{Adolphe Merkle Institute, University of Fribourg, 1700 Fribourg, Switzerland}

\affil[b]{Present address: Department of Chemistry and Physics of Materials, University of Salzburg, Jakob-Haringer-Str.\ 2a, 5020 Salzburg, Austria}

\leadauthor{Djeghdi} 

\authorcontributions{Conceptualization: K.D., U.S. and B.D.W.; Investigation - optical measurements: K.D.; Investigation - SEM imaging: K.D.; Investigation - FIB-tomography and reconstruction: K.D.; Investigation - optical simulations: K.D. and B.D.W.; Data curation: K.D.; Visualisation: K.D.; Supervision: B.D.W. and U.S.; Funding acquisition: U.S.; Writing - original draft: K.D.; Writing - Review \& Editing: all authors.\\}
\authordeclaration{The authors declare no competing interests.}

\correspondingauthor{\textsuperscript{1}To whom correspondence should be addressed. E-mail: bodo.wilts@plus.ac.at}


\begin{abstract}
\nolinenumbers
The bright colors of \textit{Pachyrhynchus} weevils originate from complex dielectric nanostructures within their elytral scales. In contrast to previous work exhibiting highly ordered single-network diamond-type photonic crystals, we here show by combining optical microscopy and spectroscopy measurements with 3D FIB tomography that the blue scales of \textit{P.\,congestus mirabilis} differ from that of an ordered diamond structure. Through the use of FIB tomography on elytral scales filled with Pt by electron beam-assisted deposition, we reveal that the red scales of this weevil possess a periodic diamond structure, while the network morphology of the blue scales exhibit diamond morphology only on the single scattering unit level with disorder on longer length scales. Full wave simulations performed on the reconstructed volumes indicate that this local order is sufficient to open a \rev{}{partial} photonic bandgap even at low dielectric constant contrast between chitin and air in the absence  of long-range or translational order. The observation of disordered and ordered photonic crystals within a single organism opens up interesting questions on the \rev{in the}{}cellular origin of coloration and studies on bio-inspired replication of angle-independent colors.
\end{abstract}

\begin{document}

\maketitle
\ifthenelse{\boolean{shortarticle}}{\ifthenelse{\boolean{singlecolumn}}{\abscontentformatted}{\abscontent}}{}
\pagenumbering{gobble} %removes page number
\nolinenumbers
Color in nature can be of pigmentary and/or structural origin, with pigmentary color resulting from the absorption of specific wavelengths of light by chemical substances and structural color resulting from the scattering of incident light by periodic nanostructures. Numerous organisms rely on structural color, either on its own or coupled with pigments, to provide crucial survival functions including aposematism, mating or camouflage \cite{cuthill_biology_2017,lee_role_2018}. The advantage of structural over pigmentary coloration is their brightness, angle-dependence (iridescence) and/or polarization-dependence, which are inaccessible by pigments alone \cite{vignolini_mirror_2012,tinbergen_spectral_2013,dalba_relative_2012}. A variety of photonic structures with increasing structural complexities have been identified in natural systems. These include low-dimensional periodic systems in the form of simple thin films or multilayers in insect wing scales and elytra \cite{stavenga_thin_2014}, diffraction gratings in flowers \cite{whitney_floral_2009} and more complex three-dimensional periodic sub-micron structures, which can for example be found in beetle, weevil and butterfly scales \cite{galusha_discovery_2008,welch_orange_2007,wilts_butterfly_2017}. These periodically ordered dielectric nanostructures are known as photonic crystals (PCs). Since the building blocks of PCs are positioned on a regular lattice, the structures possess both long\rev{}{-} and short\rev{}{-}range order. If the periodicity is on a length scale enabling interference in the visible spectrum and provided the refractive index (RI) contrast between the different phases is sufficiently high, PCs open a complete or partial bandgap resulting in structural coloration \cite{yablonovitch_inhibited_1987,martin_self_1999,joannopoulos_photonic_2008}.

\rev{On the contrary}{By contrast}, amorphous networks lack both short\rev{}{-} and long\rev{}{-}range order. In these networks, \rev{structure}{structural} parameters such as the connectivity, nearest neighbor distance, and bond angles are randomly distributed. Optical amorphous networks are highly diffusive due to the multiple scattering of light with little correlation in the phase or directionality of individual ray trajectories\rev{}{,} resulting in white coloration \cite{bohren_multiple_1987}.
Several species of \textit{Coleoptera} \rev{famously }{} rely on this principle to produce \rev{strikingly}{a striking} white `color' \cite{wilts_evolutionary-optimized_2018, vukusic_brilliant_2007, burg_liquidliquid_2019}. Between perfect order and complete disorder, \textit{quasi-ordered} structures possess local \rev{short range}{short-range} correlations and suppressed long-range fluctuations, while lacking long-range translational order. This encompasses, for instance, poly-crystals with small-sized domains, continuous random networks (CRN) with fixed valency, hyperuniform structures, and quasi-crystals \cite{barkema_high-quality_2000,zachary_hyperuniformity_2009,edagawa_photonic_2014}. In some special cases, the spatial correlations, which are on a scale comparable to optical wavelengths, are sufficient to generate \rev{pseudo-}{partial} bandgaps and hence structural coloration. The produced color is more diffuse, less saturated and angle-independent compared to colors produced by highly ordered PCs. Partially disordered bicontinuous structures are a less common mechanism to generate color in nature and observations have so far been limited to blue coloration, mainly in beetle scales and bird feathers \cite{yin_amorphous_2012,dong_optical_2011,pouya_discovery_2011, bermudez_structural_2020,shi_amorphous_2013, saranathan_structure_2012,shawkey_electron_2009,sellers_local_2017}.

So far, %\rev{}{in term of real-space imaging,} 
studies of amorphous structures \rev{}{in biological systems} have predominantly relied on 2D data collection from cross-sectional electron microscopy \rev{imaging}{} \cite{yin_amorphous_2012, dong_optical_2011,shi_amorphous_2013} \rev{}{, or small-angle x-ray scattering studies \cite{saranathan_structural_2015}}. While \rev{these data}{both methods} enable the \rev{reconstruction}{identification} of periodic morphologies, \rev{the imaging of}{an imaging approach that elucidates} the full three-dimensional structure is required to fully understand the interplay between the structure and the optical signature of non-periodic photonic materials \rev{}{\cite{shawkey_electron_2009}}.

In this work, we examine the full 3D structure in the elytral scales of the \textit{Pachyrhynchus congestus mirabilis} weevil and link it to the optical appearance of the scales. Previous work on \textit{Pachyrhynchus} weevils has discussed either single colored spots or rainbow-colored spots on jet-black elytra \cite{welch_orange_2007,wilts_literal_2018,chang_hereditary_2020}. These studies found an ordered single-network diamond structure (space group $Fd\bar{3}m$) in their scales \cite{wilts_literal_2018, chang_hereditary_2020}, with color variations resulting from changes in the lattice constant, chitin filling fraction and/or domain orientation. Through FIB-SEM 3D reconstructions of the photonic networks within the wing scales of \textit{P.\,c.\,mirabilis} and by using a novel Pt-based deposition technique, this study demonstrates the origin of color from ordered \rev{as well as}{and} seemingly amorphous PC structures. Comparing optical experiments with full wave 3D simulations of the 3D datasets combined with a 3D structural analysis allows insights into \rev{the }{insects' }coloration mechanisms \rev{of insects}{} and \rev{provide}{ provides} interesting further avenues to investigate \rev{their}{the} developmental cellular pathways \rev{}{of insect nanostructures \cite{wilts_butterfly_2017}, but also the use of amorphous diamond-based structures as photonic materials \cite{edagawa_photonic_2008}}.

\section*{Results and Discussion}
\subsection{Optical appearance and anatomy}

\begin{figure*}[b]
    \centering
    {\includegraphics[width=17.8cm]{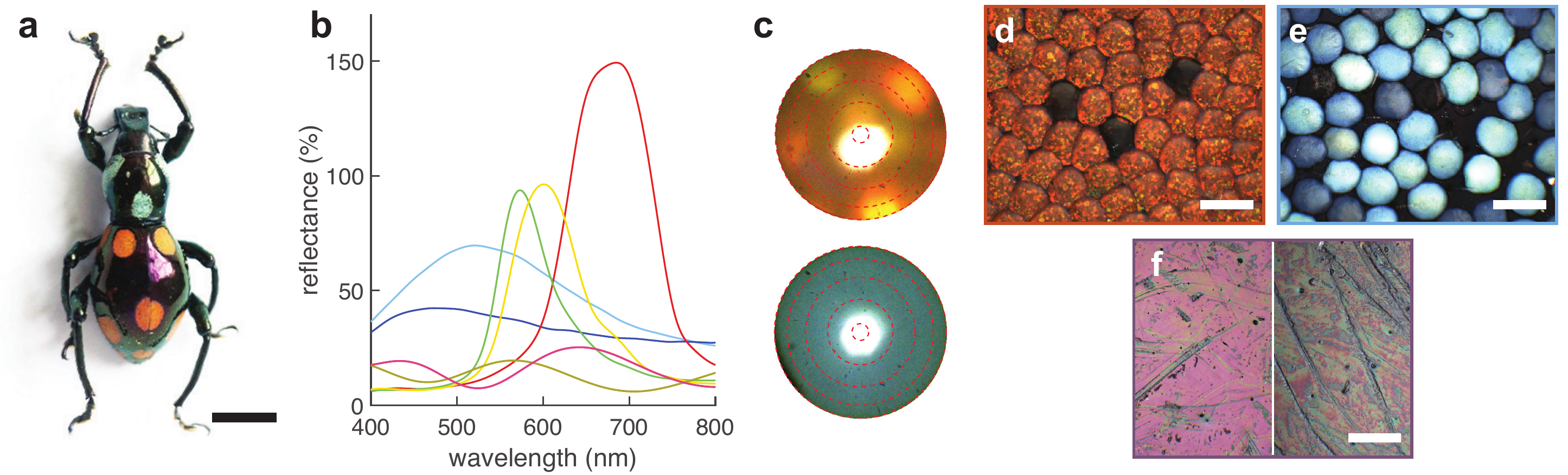}} 
    \caption{(a) Photograph of a \textit{P.\ c.\ mirabilis}. (b) Representative reflectance spectra of the elytron and scales from two different regions measured with a microspectrophotometer using a white diffuser as a reference. The blue patches contain bright and dull blue scales while red patches contain globally red scales with local yellow- and green-colored domains. The curves are plotted in a color approximately matching the observed color. (c) Scatterograms of the red (top) and blue (bottom) scales. The red-dashed circles represent scattering angles of 5\textdegree, 20\textdegree, 35\textdegree, 50\textdegree and 64\textdegree from the center to the perimeter. (d--f) Microscopy images of the various colored areas on the elytron: (d) red scales, (e) blue scales, (f) abdomen (left) and thorax (right) elytron. Scale bars: (a) 0.5\,cm, (d--f) 100\,\textmu m.}
    \label{fig:macro_spectra_scattero}
\end{figure*}

\textit{P.\ c.\ mirabilis} is an about 2\,cm long weevil with a dark-purple elytron that features red-orange and blue colored ellipsoidal spots on its dorsal and lateral sides (Fig.\ \ref{fig:macro_spectra_scattero}a). A variable number of red-orange spots are located exclusively on the abdomen
and three blue spots are located on the thorax. Each spot is assembled of brightly colored circular scales with a mean diameter of $60 \pm 2$\,\textmu m. Visual inspection at varying angles revealed a pink glare of the elytron (Fig.\ \ref{fig:macro_spectra_scattero}a,f). Although they appear uniformly colored to the naked eye, optical microscopy observations reveal that the red-orange scales contain multiple domains that vary in color (Fig.\ \ref{fig:macro_spectra_scattero}d) from green to red. In contrast, the hue of the blue scales varies, with the darker ones appearing more dull but remain uniform in color across each individual scale (Fig.\ \ref{fig:macro_spectra_scattero}e). Note that both blue and red spots also contain a few transparent scales, which are not discussed here.

To quantify the coloration and investigate its origin, we measured reflectance spectra of individual elytral scales attached to the elytron with a microspectrophotometer. Representative reflectance spectra are shown in Figure~\ref{fig:macro_spectra_scattero}b. The scales from the red-orange spots yielded narrow single peaks with full width at half maximum (FWHM) values between 70--110\,nm centered around approximately 575, 600 and 690\,nm, corresponding to green, yellow and red facets, respectively. The amplitude of the reflection peaks increased with the peak wavelength. In contrast, the dull to bright blue scales gave rise to very broad single-peaked spectra with FWHM values of approximately 220\,nm and reflectance peaks located between 470 and 515\,nm. The presence of pigments was also investigated since they are common in natural systems and are often used in combination with organismal structural colors, as observed in beetle \cite{bermudez_structural_2020} and butterfly \cite{wilts_butterfly_2017} scales or bird feathers \cite{dalba_relative_2012, tinbergen_spectral_2013}.Transmission spectra of both blue and red scales immersed in refractive index-matching oil ($n_\mathrm{o} = 1.56$, Fig.~S3) indicate the presence of a broad-band absorbing pigment that gradually increases in absorbance towards short wavelengths, reminiscent of melanin \cite{stavenga_sexual_2012}. This broad-band absorbance however does not mirror the reflectance spectra and we can therefore ascribe the coloration to a combination of structural and pigmentary origins.

In addition, reflection spectra of the elytron displayed two low-intensity reflection broad peaks characteristic of thin film interference, but the position of these peaks were different on the thorax and on the abdomen. The thorax, which bears the blue scales, appeared more black to the naked eye and displayed two peaks, one below 400\,nm ($r_{\rm{avg}} = 21\%$) and one at 565\,nm (FWHM$_{\rm{avg}} = 130$\,nm, $r_{\rm{avg}} = 16\%$). The abdomen, which carries the red scales, appeared more pink and displayed two peaks at 430 (FWHM$_{\rm{avg}} = 90$\,nm, $r_{\rm{avg}}= 18\%$) and 640\,nm (FWHM$_{\rm{avg}} = 140$\,nm, $r_{\rm{avg}} = 22\%$). 

To study the angle-dependence of the scale coloration, $k$-space measurements were performed by inserting a Bertrand lens into the imaging pathway (see Methods). Figure~\ref{fig:macro_spectra_scattero}c shows scatterograms taken with a narrow-angle illumination for each type of elytral scale. The area of illumination covered several domains of the red scales and revealed directionally-reflected spots of red, yellow and green visible at various angular positions. \rev{In contrast}{Quite differently}, the blue scales showed uniform, broad-angle scattering profiles, independent of their brightness, which are in contrast to the highly directional, narrow, high reflection peaks of the red scales. We hypothesize that these contrasting optical behaviors are the result of structural differences present within each type of scale. 

\subsection{Internal structure of the scales}
\begin{figure*}[tbp]
    \centering
    \includegraphics[width=14.6cm]{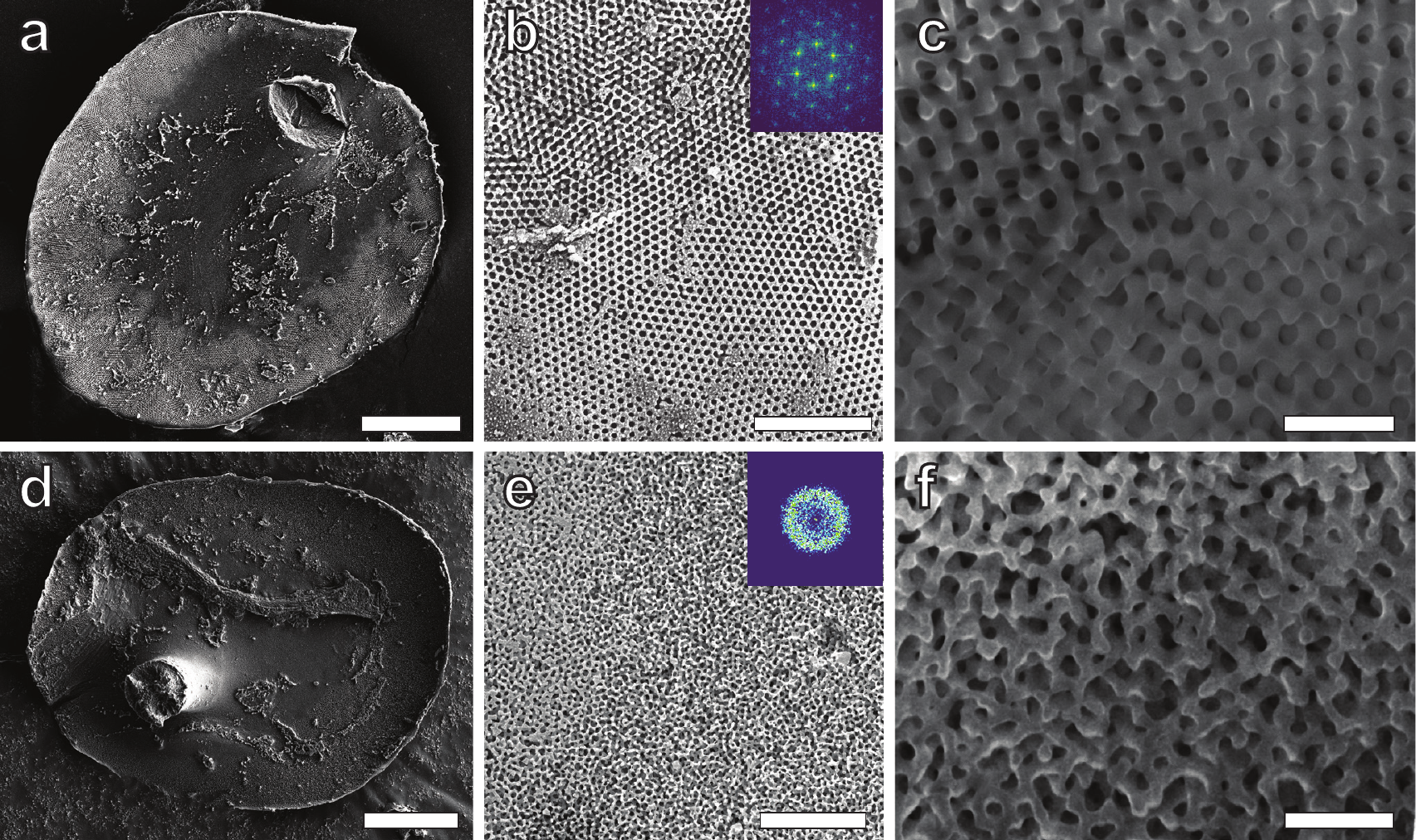}  
    \caption{SEM images of (a,d) whole scales, (b,e) top views of cortex-free regions of the scales and (c,f) cross-sectional views created by FIB milling of red (a-c) and blue scales (d-f). The red scales feature a highly organized internal structure with hexagonal symmetry, while blue scales feature an amorphous chitin structure. The insets in (b,e) show  2D fast Fourier transforms of small areas of each image to highlight local order. Scale bars: (a,d) 20\,\textmu m, (b,e) 4\,\textmu m, (c,f) 1\,\textmu m.}
    \label{fig:sem_fib}
\end{figure*}

To confirm the structural origin of the coloration and identify structural variations, top- and side-view images of each type of scale were taken using a scanning electron microscope (SEM). For the side-view cross-sectional images, scales were cut using a focused Ga-ion beam (FIB, see details in Methods). Figure~\ref{fig:sem_fib} shows the resulting overview images. 

The interior of the red scales features an ordered photonic crystal with multiple domains (Figure~\ref{fig:sem_fib}b,c). Figure~\ref{fig:sem_fib}b shows a hexagonal chitin pattern in the plane perpendicular to the normal of the scale which is characteristic of a cubic structure. The diameter of large domains reaches approx.\ 55\,\textmu $\text{m}^2$. The 2D FFT of this structure exhibits a clear six-fold symmetry with a periodicity of approximately 300\,nm (inset in Fig.~\ref{fig:sem_fib}b). Polycrystalline diamonds are common in beetles scales and have been observed through a wide range of species to generate various colors \cite{pouya_discovery_2011, galusha_discovery_2008, welch_orange_2007, ebihara_cuticle_2018}. In particular, these results are consistent with previous work on \textit{Pachyrhynchus} weevil scales, where the structure within these scales was identified as a single-diamond network of chitin and air \cite{wilts_literal_2018,chang_hereditary_2020}. The main parameters that influence the observed color of biological 3DPCs are (i) its local orientation, (ii) its lattice parameter and (iii) its solid filling fraction   \cite{joannopoulos_photonic_2008,wilts_literal_2018}. Here, the $\langle111\rangle$ orientation is clearly dominant along the surface normal, but in-plane rotations are visible, giving rise to the tessellated appearance of the scales (Fig.~\ref{fig:macro_spectra_scattero}d). Assuming a $\langle111\rangle$ orientation of a single diamond, the lattice parameter of the resulting network has a constant value of $a= 489 \pm 5$\,nm calculated from the average pore nearest neighbor distance in the $(111)$ plane, $d_{\mathrm{nn}}=346 \pm 7$ nm (Fig.~\ref{fig:sem_fib}b). From the measured chitin area fraction of the $(111)$ diamond plane, using the empirical formula $\rm{ff}_{3D}= (1.73 \, \cdot \rm{ff}_{2D,111} ) - 0.83$, the estimated 3D filling fraction is approximately 42\%. These values are in line with previously reported data on the orange-red scales of the \textit{P.\,c.\,pavonius}, which had an estimated fill fraction of 36\% and a lattice parameter of 483\ nm \cite{wilts_literal_2018}.

SEM imaging of the blue scales also shows a network of chitin in air, but no clear periodicity is visible (Fig.~\ref{fig:sem_fib}e,f). A spatial frequency was however determined by 2D FFT, which yielded a diffuse disk with a pronounced ring indicating a preferred nearest neighbor distance which is spatially not very well defined (Fig.~\ref{fig:sem_fib}e, inset). This length corresponds to the average distance between two pores of approximately 250\,nm. Since all blue scales, whether dull or bright, display similar spectral and scattering behaviors, and since observations through \rev{electronic}{an electron} microscope showed no clear variation from one blue scale to another, the remainder of the study focuses on a single representative blue scale.

\subsection{A novel 3D imaging approach using FIB-SEM}
3D data is needed to unambiguously characterize the scale structures and relate these to their photonic properties. So far, data of biophotonic structures \rev{}{collected by SEM or TEM imaging} were often limited to 2D \rev{,}{information}   \cite{dong_optical_2011,shi_amorphous_2013,yin_amorphous_2012,welch_orange_2007, wilts_brilliant_2012, bermudez_structural_2020}. Moreover, the process of obtaining a single cross-section from a biological sample through TEM involves lengthy sample preparation \rev{, consisting}{that consists} of embedding, staining and lift-out, sometimes followed by demanding post-processing to deconvolute the resulting image \cite{vukusic_physical_2009, fleck_fib-sem_2019}. These methods are often `blind' so that a desired location within a 3D morphology cannot be extracted. 

\rev{}{Numerous studies have used small-angle x-ray scattering (SAXS) to gather 3D information from a wide range of biophotonic samples \cite{saranathan_structure_2010,saranathan_structural_2015,saranathan_structure_2012}. For ordered systems, SAXS often enables space group assignment as well as quantification of structural parameters. The SAXS results for disordered structures are less clear and show a diffuse disk for amorphous structures, while quasi-ordered structures feature a pronounced ring from which short-range spatial periodicity can be extracted. The assignment of a full 3D morphology using SAXS is however not always unambiguous and can benefit from complementary real-space imaging techniques. Imaging is especially beneficial in the case of amorphous and quasi-ordered system, as it allows direct visualization of the 3D arrangement of matter in a chosen region of interest, while SAXS usually averages over a larger volume and does not give information on local arrangements.} 3D \rev{data}{imaging} on this length scale (10\,nm -- 1 \textmu m) can \rev{however}{also} be achieved through ptychographic X-ray-based tomography \cite{wilts_evolutionary-optimized_2018}. \rev{,which}{Both SAXS and X-ray tomography however} require access to a synchrotron \rev{}{with distinct beamline configurations} or specialized lab equipment \cite{wilts_butterfly_2017}. \rev{In contrast,}{Real-space tomography techniques are a good alternative, complementary technique to consider. In particular,} FIB-SEM tomography datasets can be automatically acquired in a few hours and require very little ex-situ preparation, while allowing for the extraction of a very large set of cross-sections from any direction. FIB-tomography has been previously described for a few biological samples \cite{galusha_discovery_2008, ebihara_cuticle_2018}, but the analysis was limited to a qualitative space group assignment. Perhaps constrained by the reconstruction process, no in-depth structural characterization was performed on the reconstructed volumes. The main limiting factor is not the inherent instrument resolution, but lies in the sample preparation. A backfilling step of the air-cuticle material is necessary for any 3D reconstruction resulting from a slice-and-view process to limit the depth of field and to guarantee a sufficient contrast for accurate automated binarization. This is usually achieved through filling of the sample with a low viscosity resin prior to tomography \cite{galusha_study_2010}, a time-consuming ex-situ process that is often ill-controlled and ultimately yields a poor image contrast. 

As part of this study, a novel approach based on in-situ CVD-based filling inspired by Eswara-Moorty \textit{et al}.\ \cite{eswara-moorthy_situ_2014} was developed. Platinum (Pt) was deposited into the porous structure using an electron-beam induced deposition process (EBID). A precursor gas containing Pt was injected close to the ROI and decomposed into the air phase by interacting with a secondary electron beam (for details, see Methods). Choosing Pt backfilling was primarily motivated by the ease of the in-situ process, allowing precise control over the area and depth of deposition in the electron microscope. This backfilling tremendously benefited image acquisition and analysis due to the enhanced contrast between the Pt and the chitin network, as well as reducing curtaining effects and charge accumulation (see Fig.~S1). Another advantage was the preservation of the integrity of the structure by protection against beam-induced melting and deformation. The combination of all these parameters enables a great fidelity of the reconstruction and better reliability of the structural analysis compared to other methods.

\begin{figure*}[t]
    \centering
    \includegraphics[width=16.6cm]{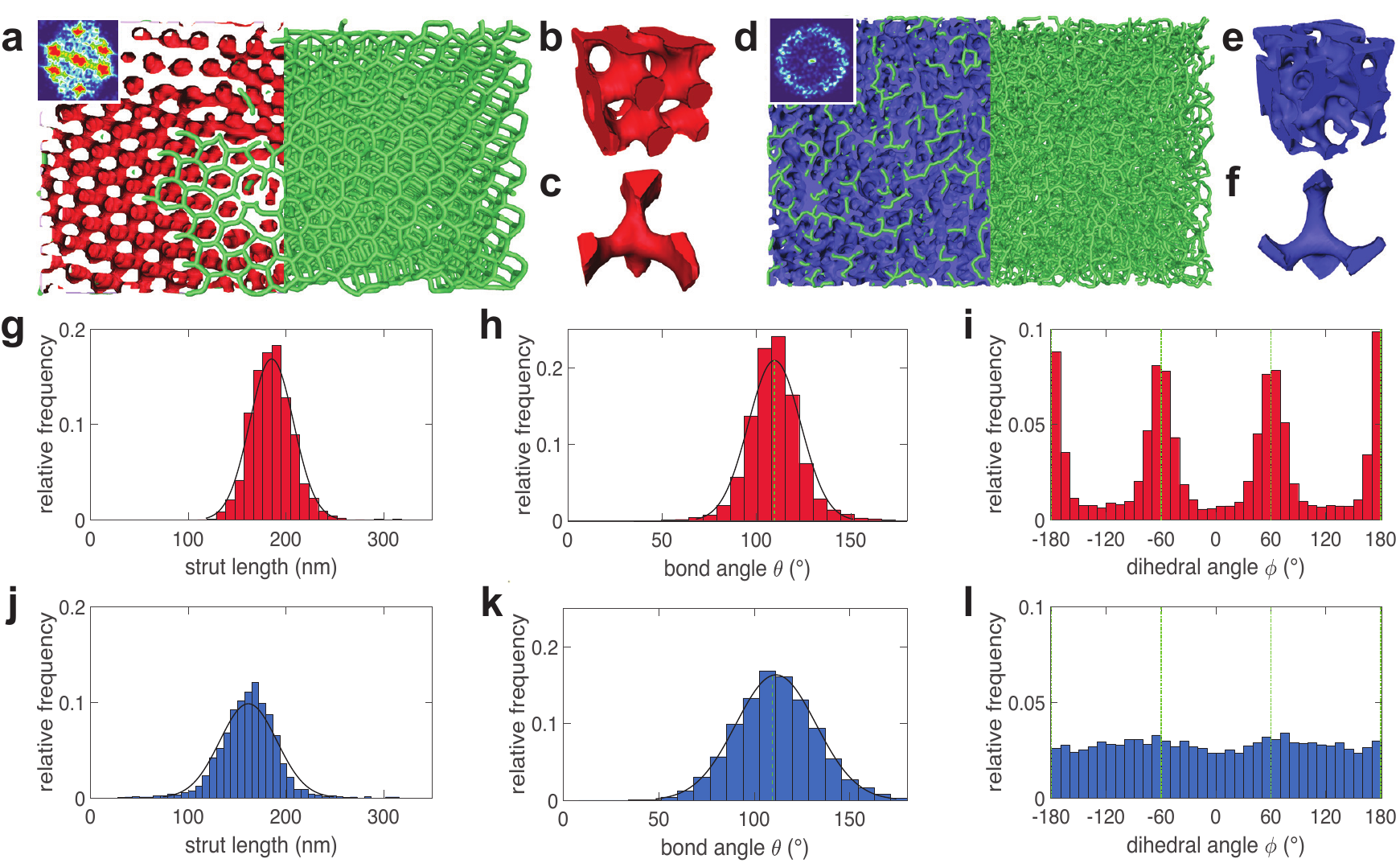}
    \caption{(a,d) Full 3D reconstruction (left) and skeletonization (right) of the chitinous structures present in the red and blue scales, $V = 5 \times 4 \times 3$\,\textmu $\text{m}^3$. Both structures are seen from the top \textit{i.e}.\ the plane corresponds to $x-z$. Insets: cross-section of the respective 3D FFTs. (b, e) Small volume $V \approx (550\,\text{nm})^3$ and (c, f) single node reconstruction. (g, j) Strut length distributions for similar strut counts ($N_{\rm{red}} = 5208$, $N_{\rm{blue}} = 3494$). (h, k) Bond angle distributions ($N_{\rm{red}} = 18571$ , $N_{\rm{blue}} = 21139$). (i, l) Dihedral angle distributions ($N_{\rm{red}} = 28568$, $N_{\rm{blue}} = 22014$). (a-c, g-i): red scale, (d-f, j-l): blue scale.}
    \label{fig:Avizo}
\end{figure*}
 
\subsection{3D volumetric imaging of weevil scales} 
Using this novel backfilling procedure, FIB-tomography was performed using a slice-and-view process and the obtained image stacks were converted into 3D reconstructions (see Methods for details). Note that since the cortex was removed prior to the tomography, only the structure within the scale was imaged. In brief, the main workflow to reconstruct 3D structures consisted of a registration step, followed by cropping, filtering, and segmentation for reconstruction. The reconstruction resulted in two data-sets of $5 \times 4 \times 3$ \textmu m$^3$ with a voxel resolution of 10\,nm edge length. This enabled the visualization and structural analysis of sizable volumes (Fig.~\ref{fig:Avizo}a,d), the extraction of supercells (Fig.~\ref{fig:Avizo}b,e) and single repeat units (Fig.~\ref{fig:Avizo}c,f). Additional visuals are shown in \rev{the supplementary information (Fig.\S4a--d)}{Figure~S4a--d}. The chitin filling fraction estimated from the binary voxelized volume was 42\% for the red scale, matching the estimation from the 2D area fraction, and 32\% for the blue scale. Note that a precise determination of filling fraction is challenging as it is highly dependent on the threshold and the algorithm used during the binarization process. The values obtained here are in the range of previously reported values for related \textit{Pachyrhynchus} \cite{welch_orange_2007, wilts_literal_2018} and \textit{Entimus} \cite{wilts_brilliant_2012} weevils, and are not far from the optimal filling fraction that maximizes the frequency width of the optical bandgap of a single-diamond PC \cite{wilts_brilliant_2012}. 

The full reconstructed volume of the red scale comprised multiple domains with recognizable patterns (Fig.~S4a,b), although the top plane remained in the (111) orientation throughout. The unit cell appeared as a well-defined structure made of regularly connected tetrahedral units (Fig.~\ref{fig:Avizo}b,c). The unit cell was identical throughout the entire volume, as long as it was extracted away from a grain boundary. 

The structure within the blue scale exhibited no visible grain boundaries and the unit cell appeared irregular and tortuous (Fig.\ \ref{fig:Avizo}d,e and Fig.\ S4c,d). Tetrahedral arrangements of nodes were also observed (Fig.\ \ref{fig:Avizo}f) but the connectivity was less uniform compared to the ordered structure of the red scales. In particular, a number of trigonal planar nodes arrangements were present, decreasing the average coordination number. Cross-sectioning the 3D FFT of each structure yielded similar patterns to the one observed in 2D, \textit{i.e}.\ a hexagonal pattern for the ordered structure and a well\rev{}{-}defined ring for the amorphous one (insets in Fig.~\ref{fig:Avizo}a,d), indicative of a full 3D isotropy.

To fully characterize the diamond structure of the red scales and to investigate a possible underlying order in the blue scales, we performed a skeletonization step followed by a morphological study. Based on the skeleton of each structure, the average coordination number of each vertex (or node) was extracted as well as the distributions of the strut lengths, bond angles between connected segments and dihedral (or torsion) angles between intersecting planes (Fig.\,\ref{fig:Avizo}g--l). The reconstructed red scale possessed an average strut length $L_\mathrm{r}=185 \pm 22$\,nm ($N=5208$), an average coordination number $CN_{\rm{r}}=3.8 \pm 0.4$ ($N = 3949$) and an average bond angle $\theta_{\rm{r}}=109.6 \pm 13.9$\textdegree\, ($N=18571$). The same study conducted on a blue scale over a comparable subvolume yielded $L_\mathrm{b}=162 \pm 29$\,nm ($N=3494$), $CN_{\rm{b}}=3.4 \pm 0.6$ ($N =9983$), $\theta_{\rm{b}}=111.3 \pm 21.0$\textdegree\, ($N=21139$). 

\begin{figure*}[tbp]
    \centering
    \includegraphics[width=14.6cm]{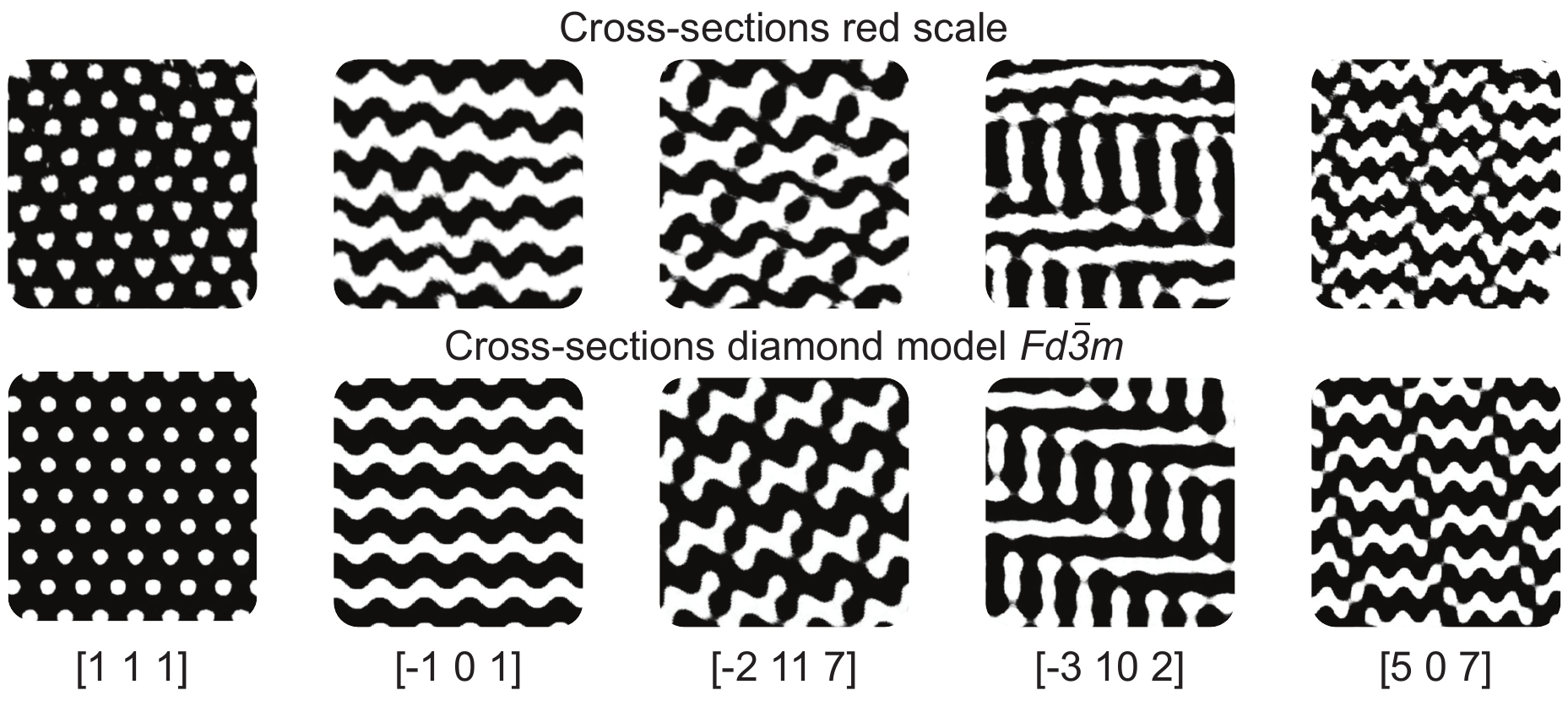}   
    \caption{Matching cross-sections between the tomogram of the red scale (top row) and the model diamond (bottom row). Chitin is shown in white, air in black.}
     \label{fig:CS}
\end{figure*}

In both cases, the values of strut lengths and bond angles followed a Gaussian distribution, with the disordered network having a slightly higher standard deviation. The statistical analysis of the ordered network yielded values of node connectivity and bond angles close to the values expected for the ideal diamond \textit{i.e}.\ $CN_\mathrm{d} = 4$ and $\theta_\mathrm{d} = 109.5$\textdegree. The disordered network showed a slightly lower average connectivity but similar average bond angles. Note that the angle distribution is a more robust parameter than the connectivity as it is less prone to reconstruction artifacts. Because of binarization and skeletonization artifacts, there is a tendency to underestimate the coordination number and since the average bond angle is close to the ideal tetrahedral angle, it seems that the node positions of the disordered network are consistent with a tetrahedral distribution. The dihedral angle distribution of both scales showed local maxima around the values expected for a diamond \textit{i.e}.\ $\phi_{d} \in \{\pm 180;\pm 60\} $\textdegree, but while the ordered network yielded well-defined, narrow peaks, the disordered network shows a nearly constant angular distribution with only a minor undulation. The blue scale thus appears much more tortuous than the red scale at medium range (Fig.\,\ref{fig:Avizo}b,e), which is a result of about random torsion angles between tetrahedral units.

Some part of the distribution width of the measured quantities can be attributed to the data acquisition and reconstruction processes, combined with topological and positional defects accounting for grain boundaries and other inherent imperfections, and defects from the biological processes that generated the network structures \cite{sellers_local_2017, mouchet_optical_2020,tsimring_noise_2014}.%johnston_mitochondrial_2012,
Since these contributions should be identical for both systems, the wider spread observed for all measured quantities in the blue scale can be ascribed to a higher `disorder'. The difference is most notable in the dihedral angle distribution, which is coherent with decreasing structural correlations on  medium to long length scales. 

The sizable reconstructed volume also allows \rev{to extract}{extracting} TEM-like cross-sections that can be compared to cross-sections of known space groups generated by level-set equations. A qualitative analysis of these cross-sections can already shed light on the nature of the networks and reveal any anisotropy. This method can be used in combination with  methods to attribute unknown structures to a space group \cite{wilts_brilliant_2012, michielsen_gyroid_2008}. To further confirm the $Fd\bar{3}m$ space group of the red structure, planes exhibiting a recognizable pattern were extracted from the volume and compared to the ideal diamond lattice generated from the level-set equation. To this end, the computed model was rotated and cross-sectioned incrementally until planes matching the experimental data were found. Figure~\ref{fig:CS} shows compelling resemblances between the cross-sections of the red scales and the single-diamond model, confirming the space group $Fd\bar{3}m$ of the red scale. Other 3D network geometries were similarly investigated, such as the single-network gyroid structure (space group $I4_132$), which is also often found in insects \cite{wilts_butterfly_2017, saranathan_structural_2015}, but no matching cross-sections were consistently found. 

\begin{figure}[htb]
    \centering
    \includegraphics[width=0.75\linewidth]{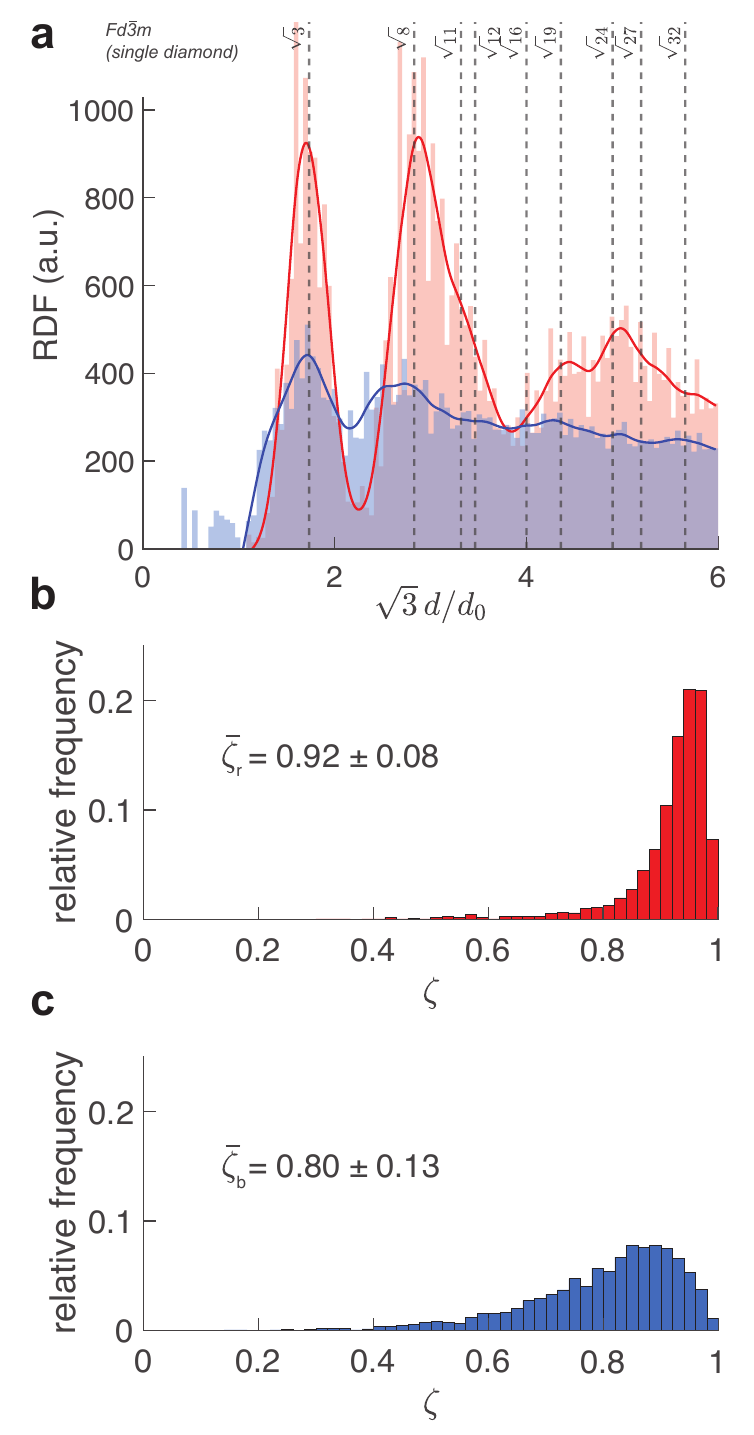}  
 \caption{(a) Radial distribution function computed for two scales. The abscissa is normalized by $d_0/\sqrt{3}$ where $d_0$ is the average strut length. Histogram distributions of the scales are shown in translucent red and blue to match the color of the scales they correspond to. Cubic splines are shown as opaque lines of the same color. The peak positions of the single diamond $Fd\bar{3}m$ are indicated as vertical dashed lines. (b) Distribution of the tetrahedral order parameter $\zeta$ calculated from a subvolume of the red scale ($N_{\rm{nodes}} = 2406$) and (c) the blue scale ($N_{\rm{nodes}} = 3161$).}
\label{fig:RDF}
\end{figure}

The same procedure was carried out for the reconstructed internal structure of the blue scale and yielded similar-looking cross-sections regardless of orientation, which is consistent with the apparent isotropy of the network. Observations made along the main directions ([001], [100], [010], [110], [111]) are shown in Fig.~S4e. Since no recognizable pattern appeared, this scale obviously does not possess a well-ordered 3D symmetric structure.

\subsection{Volumetric structural order}
The main distinction between an ordered and a quasi-ordered structure is that the latter only possesses short range order \textit{i.e}.\ local order of the scattering centers gives way to disorder on longer length scales. Radial distribution functions (RDF) of the vertex positions in each sample \cite{zallen_physics_1998} were computed to estimate the distance over which local order disappears. Figure~\ref{fig:RDF}a shows the calculated RDF for the two obtained volumes.

The red sample displays clear local peaks expected for a diamond structure with maxima at $\sqrt{3}, \sqrt{8}, \sqrt{19}$ and $\sqrt{24}$ of the scaled distance $d/d_0$, calculated from the reflection conditions of the $Fd\bar{3}m$ space group \cite{hahn_international_2005}. The RDF of the blue scales, however, shows a broad background and peaks that are much less pronounced compared to those of the red scale, characteristic of a broad distribution of nearest neighbor distances. Nevertheless, the blue scale clearly displays the first two $Fd\bar{3}m$ peaks, revealing short-range order and a structural similarity to a diamond network. The absence of higher-order peaks attests the lack of long-range order, similar to amorphous solids \cite{zallen_physics_1998,jin_photonic_2001}.

To further quantify the degree of short-range order on the scale of single repeat units, a tetrahedral order parameter was calculated for every node with a fourfold connectivity (Fig.~\ref{fig:RDF}b,c). %A rescaled version \cite{errington_relationship_2001} 
For this, we used a rescaled version of the order parameter introduced by Chau \textit{et al}.\ \cite{chau_new_1998} via
\begin{equation}
    \zeta= 1-\frac{3}{8} \sum_{j=1}^3  \sum_{k=j+1}^4 \left(\cos\, \theta_{jk}+ \frac{1}{3}\right)^2,
\end{equation}
where $\theta_{jk}$ is the bond angle between each edge connected to the same node. In a four-connected structure with randomly oriented edges, $\bar{\zeta}_\mathrm{rand}\,=0$. In a perfectly periodic diamond network, all bond angles are  $\cos^{-1}(\theta_{jk})=- \frac{1}{3}$ such that $\bar{\zeta}_{d}=\zeta_{d}=1$ for all $j,k$ values. An analysis of the reconstructed scales yielded average tetrahedral order parameters of $\bar{\zeta}_\mathrm{r} = 0.92 \pm 0.08$ for the red scale and $\bar{\zeta}_\mathrm{r} = 0.80 \pm 0.13$ for the blue scale, which is consistent with their respective bond angle distributions. The shape of the distributions, with only one local maximum centered around a value close to 1, indicates that both networks have quite uniform local topologies across the entire investigated volumes similar to that of an ideal diamond lattice \cite{liew_short-range_2011}.

\subsection{Optical modeling and bandgap criteria}
\begin{SCfigure*}[\sidecaptionrelwidth][tbp]
     \centering
     {\includegraphics[width=12cm]{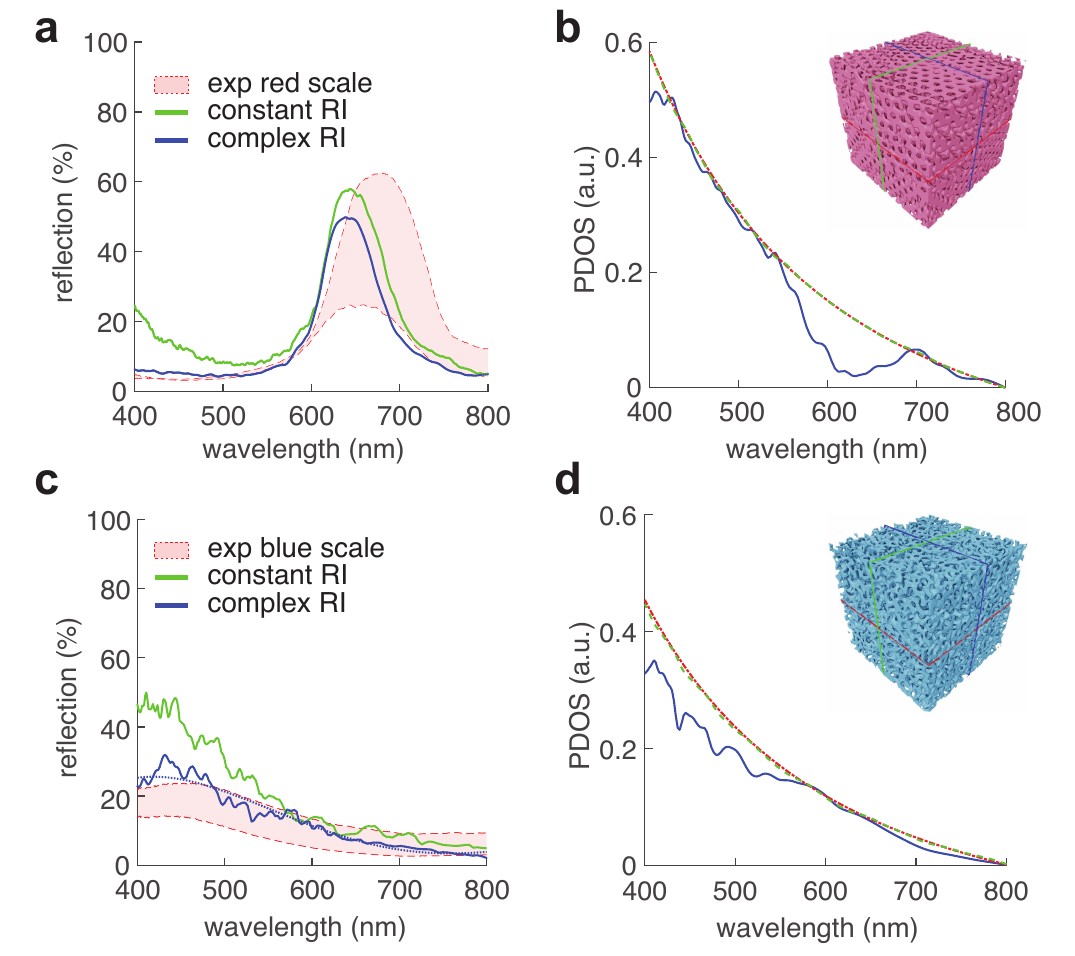}}
     \caption{Simulated reflection spectra based on the 3D tomogram of a red scale (a) and of a blue scale (c) with the refractive index dispersion of unpigmented chitin (green line) and pigmented chitin (blue line). The experimental spectra are shown as red shaded areas. Photonic density of states for the red (b) and blue (d) scales were calculated for the unpigmented chitin, plotted in blue. The  theoretical and simulated PDOS for an equivalent homogeneous medium are plotted in red and green, respectively. Insets: 3D volumes used as input for the FDTD simulations.}
 \label{fig:FDTD}
\end{SCfigure*}

To understand the optical function of both networks and validate our reconstruction process, FDTD calculations based on the acquired volumetric data were performed. Simulations resulted in the reflection spectra shown in Figure~\ref{fig:FDTD}a,c, which were compared with experimental data (red-shaded bands show the experimental variations). 

In a first approximation, the refractive index of chitin was set to $n_\mathrm{c}=1.56$ for the entire spectrum (green lines in Fig.\ \ref{fig:FDTD}a,c). These simulation results deviated significantly from the experimental data at wavelengths below 500\,nm. The assumption of a wavelength-dependent dispersion of the complex refractive index, accounting for both the wavelength dependence of the chitin \cite{stavenga_thin_2014} as well as the absorption coefficient of the pigment within the chitin (for details, see Methods), produced simulated spectra that improved the match with the experimental ones, both in terms of peak wavelength and overall reflection intensity particularly at short wavelengths where pigment absorption becomes significant (blue lines in Fig.\ \ref{fig:FDTD}a,c). The blue shift of the simulated spectra compared to the directly measured data may arise from a different scale location or from a slight compression of the scales during processing of the volumetric data. 

A further important aspect is the presence of a pigment in both types of scales. The role of the broad-band absorbing pigment is primarily to suppress short wavelength reflections and reduce multiple scattering - as previously observed in other animals and manufactured systems \cite{forster_biomimetic_2010, jacucci_limitations_2020, hu_amorphous_2019}. In the case of the red scales, the pigment is crucial in suppressing incoherently scattered blue light from the defects in the scale structure. Without the pigment, the scale would have an unsaturated, more purplish appearance (Fig.~\ref{fig:FDTD}a, green line). Blue structural coloration is possible without the pigment contribution (Fig.~\ref{fig:FDTD}c, green line), but also here, the pigment suppresses scattered stray light and stabilises the hue. Overall, the good match of the simulated and measured reflection spectra in Figure~\ref{fig:FDTD}a,c, using the reconstructed scales as an input, validates our tomography and reconstruction protocol, which should be readily expandable to other porous media with similar internal length scales.

To determine the physical origin of this coloration, the photonic densities of states (PDOS) were calculated for each scale, shown in Fig.~{\ref{fig:FDTD}b,d} (blue lines). In this calculation, a constant refractive index of 1.56 was used, i.e.,\ only the structural contribution was taken into account. These were compared to the theoretical and simulated PDOS of equivalent homogeneous media (Fig.\ \ref{fig:FDTD}b,d, red and green lines, respectively). In these equivalent media, the RI was taken as a filling fraction-weighted average of the refractive indices of chitin and air. For both cases, the PDOS determined from the scale networks show significant deviations from the expected quadratic behavior \cite{novotny_principles_2006}. The ordered structure displays a clear dip between 545--695\,nm, which corresponds to the position of the simulated reflection peak. The dip almost reaches zero and possesses a gap width to mid-gap frequency ratio of 24\%. The deviation of the disordered structure is more subtle as no localized dip is observed in the simulated range. Nevertheless, at low wavelengths, the PDOS values are significantly lowered with respect to the homogeneous medium, giving rise to a broad \rev{pseudo-}{partial band}gap ranging from 400 to 595\,nm, in accordance with the simulated reflection peak of the corresponding structure.  

Although \rev{the}{a} bandgap is expected for a `perfect' diamond, the \rev{}{partial} bandgap of the ordered structure shows a high resilience of the optical properties with respect to structural imperfections, such as grain boundaries, imperfect periodicity, positional and topological defects \cite{mouchet_optical_2020}. In an amorphous medium, the incoming light undergoes strong isotropic scattering usually resulting in white coloration. The presence of a bandgap requires a sufficient RI contrast between the two phases and local structural correlations on at least a local scale to enable a wavelength-selective reflection process through coherent scattering \cite{yin_amorphous_2012, prum_coherent_1998}, and%, shi_macroporous_2010}. 
long-range fluctuations need to be minimized to inhibit diffuse scattering \cite{man_isotropic_2013}. The investigation of amorphous 3D network structures in biological systems leading to \rev{an}{a partial} optical bandgap has recently gained increased traction. For instance, investigated structures resulting from spinodal decomposition or random close packing of spheres are known to generate colors (often blue) in bird feathers and beetle scales \cite{shawkey_electron_2009, saranathan_structure_2012,shi_amorphous_2013}. Recently, model calculations by Edagawa and colleagues introduced an amorphous network exhibiting a complete bandgap, the so-called photonic amorphous diamond (PAD), a tetrahedral network with only short-range order, as opposed to the photonic crystalline diamond (PCD) which also possesses long-range and translational order \cite{edagawa_photonic_2008}. Surprisingly, the PAD bandgap was of almost the same quality as the PCD. Sellers \textit{et al}.\ showed that the formation of a bandgap is favored by connected networks displaying low (< 5) and uniform coordination numbers. Amorphous structures with trigonal and tetragonal strut connections (connectivities of three and four, respectively) are champion structures for isotropic \rev{pseudo}{partial} bandgaps, due to the high symmetry of their scattering centers  \cite{liew_photonic_2011, sellers_local_2017, weaire_existence_1971}. 

The criterion for symmetry is the ability to superimpose scattering centers upon rotation and permutation, which requires, in addition to a uniform connectivity, a narrow distribution of bond angles. Our results corroborate this hypothesis by comparing the networks of the two scale types, which i) possess a relatively uniform connectivity that lies between three and four, ii) possess scattering centers with high symmetry, with values of the tetrahedral order parameter reaching 0.80 and 0.92. Thirdly, the less periodic structure of the blue scale, which displays a wider spread in coordination number and bond angles,  exhibits a PBG which is less localized and intense compared to the periodic diamond morphology of \rev{}{the }red scale (Fig.\ref{fig:FDTD}b,d). 

\section*{Conclusion}

The investigated specimen is set apart from other \textit{Pachyrhynchys} weevils by its display of two separate sets of colored scales, each comprising a structure with a different degree of order. While the coexistence of both ordered and disordered inner structures on a single specimen has previously been reported in the scales of other beetles (e.g.\ \textit{Eupholus magnificus} \cite{pouya_discovery_2011}, \textit{Sulawesiella rafaelae} \cite{bermudez_structural_2020}\rev{}{, and \textit{Sternotomis pulchra bifasciata} \cite{saranathan_structural_2015}}), these studies have \rev{}{either} mostly focused on their optical properties with structural information being inferred from 2D cross-sections only \rev{}{or were limited in their characterization to structure diagnosis}. An in-depth characterization of the network structures is \rev{}{however} critical not only to understand the optical properties, but also the underlying cellular processes leading to their formation, which could inform new approaches to the design and manufacture of bio-inspired optical devices \cite{vaz_photonics_2020, tadepalli_bio_2017}. Although the precise intracellular mechanism of scale formation in beetles is unknown, it is possibly homologous to the growth of butterfly scales that form as a result of the co-assembly of an infolding lipid-bilayer membrane and cuticle in the lumen of the developing scale  \cite{wilts_butterfly_2017,seago_evolution_2019}. 

The present work allows the first direct comparison of structural coloration arising from a crystalline structure and its amorphous counterpart, supported by an optical study, 3D structural data analysis as well as simulations. The 3D analyses show that structural differences lie at the heart of the difference in spectral and spatial dependency between the red and blue scales. Local order was observed in both structure\rev{}{s}, with a fixed strut length and a narrow distribution of angles leading to a high value of their respective tetrahedral order parameter. However, the circular pattern of the FFT, the shape of the RDF and the wide distribution of dihedral angles for the blue scale clearly show a lack of long\rev{}{-}range order compared to the ordered red scale, and is characteristic of the quasi-ordered nature of the network. Diamond structures are known to open an optical bandgap both in their crystalline (PCD) and amorphous form (PAD) \cite{edagawa_photonic_2008}, yet most studies on PADs have been purely computational and yielded substantial bandgaps only for high dielectric contrast (Si-air, \textit{i.e}.\ 13:1). Our results show an experimental sample displaying both a PCD and a PAD capable of generating color at low dielectric contrasts (chitin-air \textit{i.e}.\ 2.4:1). An optical bandgap, even if only partial, in a low dielectric contrast quasi-ordered material is remarkable and carries significance for the design and fabrication of quasi-ordered PC.

\matmethods{
\subsection*{Specimen}
\textit{Pachyrhynchus congestus mirabilis} weevils (Billberg 1820; Curculionidae: Entiminae: Pachyrhynchini) native to the Philippines were purchased from thebugmaniac.com.

\subsection*{Optical Characterization}
Spectral characterization was performed using a xenon light source (Thorlabs SLS401; Thorlabs GmbH, Dachau, Germany) and a ZEISS Axio Scope.A1 microscope (Zeiss AG, Oberkochen, Germany). The light reflected from the sample was collected by an optical fiber (230\,\textmu m core) in a plane confocal to the image plane, resulting in an effective measurement spot diameter of $\sim$13\,\textmu m at a magnification of $\times20$. The spectra were analyzed by a spectrometer (Ocean Optics Maya2000 Pro; Ocean Optics, Dunedin, FL, USA). Optical micrographs were captured with a CCD camera (GS3-U3-28S5C-C, FLIR Integrated Imaging Solutions Inc., Richmond, Canada). Reflection spectra were taken on scales still attached to the elytron. Transmission spectra were also taken with the same equipment, but the scales were first detached from the elytron and deposited on a microscope glass slide using a scalpel, before immersion in a refractive index matching oil ($n_\mathrm{o}=1.55$). Scatterometry measurements were performed by placing a Bertrand lens (Zeiss 453671) into the imaging pathway and using a high numerical aperture air objective (Zeiss Epiplan Neofluar 100x, NA 0.9). This allowed measurements of scattering angles up to 64\textdegree. 

\subsection*{Electron microscopy and tomography}
Individual scales were gently scratched from the elytron and deposited on an aluminum SEM stub (Plano-EM, Wetzlar, Germany) covered with conductive carbon tape. The cortex layer surrounding individual scales was partially removed by plasma etching using a PE-100-RIE system (Plasma Etch Inc., Nevada, USA). Samples were exposed to a 4:20 \ce{O2}/Ar plasma mixture for 12 to 18 minutes. The stub was then sputter coated with a 7\,nm thick layer of either gold or platinum using a Cressington 208 HR (Cressington Scientific Instruments, Watford, England). Copper tape and silver paste were also added to increase the conductivity of the sample and limit charging effects. Top-view scanning electron microscope (SEM) pictures were taken using a Tescan Mira3 (Tescan, Brno, Czechia) with a beam voltage of 8\,kV and a working distance of 10\,mm. Cross-sections were milled and imaged using a Thermo Scientific Scios 2 DualBeam FIB-SEM (FEI, Eindhoven, the Netherlands).

To enable 3D reconstruction, the scales were first filled \textit{in situ} with platinum using an electron beam-induced deposition (Pt-EBID) process inspired by Eswara-Moorthy \textit{et al}.\ \cite{eswara-moorthy_situ_2014}. A gaseous precursor, \ce{C9H16Pt}, was injected near the surface of the cortex-free region of interest (ROI) through the gas-injection system needle, and dissociated into Pt by interacting with the electron beam set to an acceleration voltage of 30\,kV, a current of 1.6\,nA and a dwell time of 15\,\textmu s. These parameters were optimized to enable a complete infiltration throughout the entire scale thickness. A rough cut was milled in front of the ROI to expose the imaging plane and trenches were milled on the sides to provide a deposition site for the milled material and prevent redeposition on the exposed section. A fiducial was created to limit beam-induced drift, stage displacement and tilt. The tomography process was automatized using ThermoFisher Auto Slice and View software (v.\ 4). Slices of 10 to 30\,nm thickness were milled with the \ce{Ga+} beam set at 30\,kV acceleration voltage and 0.30\,nA beam current. Images were acquired after each slice in the OptiTilt configuration using the built-in SEM Everhart-Thornley (ETD, secondary electrons) and in-lens T1 (A+B composite mode, back-scattered electrons) detectors set to a voltage of 2\,kV. The image distortion induced by the acquisition at a 52\textdegree\ angle was compensated with the built-in tilt correction feature.

\subsection*{Reconstruction}
Stacks of images were processed using Fiji \cite{schindelin_fiji_2012} and FEI Avizo\textsuperscript{\texttrademark} for Materials Science 2020.2 software for basic image processing, 3D reconstruction and statistical analysis. Registration was performed using a combination of the Fiji \textit{StackReg} \cite{thevenaz_pyramid_1998} and \textit{Correct 3D Drift} plug-ins \cite{parslow_sample_2014}.
A median filter was applied to despeckle and smooth the images, which were then inverted to make the chitinous material appear white and the Pt filling (corresponding to the air network) black. Segmentation was done using the trainable Weka segmentation 3D plug-in \cite{arganda-carreras_trainable_2017} in Fiji. The images were sampled down to yield isotropic voxels and to limit artifacts. Skeletonization was then performed in Avizo using an algorithm based on distance mapping and thinning. The study of the structure revealed anisotropy in the direction normal to the scale which was found in all scales of the \textit{Pachyrhynchus} family and systematically led to a bimodal strut length distribution. As previous small\rev{}{-}angle x-ray scattering studies did not reveal this anisotropy \cite{wilts_literal_2018, saranathan_structural_2015}, we believe this to arise from the FIB-SEM imaging process. The images were thus corrected by applying a vertical stretch of about 1.2. The implementation of this step lead to isotropic segment lengths, in line with previously reported data. The strut length distribution, average coordination number (CN) and filling fraction were computed using Avizo built-in features. The strut lengths were only computed on struts connected to nodes with \rev{a}{} fourfold connectivity \rev{in order }{}to limit artifacts due to struts with uneven radii that are disconnected by the skeletonization algorithm. Bond and dihedral angles were computed in Matlab v.2019b from the positions of all connected nodes derived from the skeleton (see \rev{SI }{}Fig.~S2).

\subsection*{Optical Modeling} \label{section: modeling}
The optical properties of an idealized single network diamond (space group $Fd\bar{3}m$), approximated by Schwarz's D triply periodic minimal surface model, was generated from its level-set equation following the IMDS (inter-material dividing space) method \cite{wohlgemuth_triply_2001,michielsen_gyroid_2008}, $ \cos(2 \pi z /a) ~ \sin(2 \pi (x+y)/a) + \sin(2 \pi z/a) ~ \cos(2 \pi (x-y)/a) = \ t
\label{eq:diamond} $, where $a$ is the cubic lattice constant and $t$ is the level-set parameter that determines the chitin filling fraction $v_\text{f}$ by the relation $t=2.4 \, (v_\text{f}-0.5)$.
The models were generated using Matlab with a unit cell \textit{a} chosen to match the lattice parameter determined from the SEM images.

Finite-Difference Time-Domain (FDTD) calculations \rev{based on the 3D reconstructions generated from the measured tomograms}{that directly used the obtained 3D reconstructions as input} were performed using the Lumerical software (Lumerical FDTD Solutions, v. 8.20; Ansys Inc, Canonsburg, PA, USA) to predict reflection spectra and photonic density of states (PDOS). The refractive index of the material was chosen to be a complex function of the wavelength to take into account both the effects of the chitin and the absorbing pigment present in the scales. Since the pigment contribution is mainly absorptive, the real part was taken equal to the chitin RI which is well approximated by a wavelength-dependent Cauchy law (see SI, eq.\ 1) \cite{stavenga_sexual_2012}. The imaginary part of the RI (the absorption coefficient) was calculated from the experimental transmission spectra taken in a refractive index matching fluid (see SI, eq.\ 2 and Fig.~S3).

For local PDOS calculations, three dipoles, each oriented along one direction ($x$, $y$, $z$), were placed in proximity of the structure and the electrical field inside the structure was calculated using a transmission box surrounding it. The total local density of states was subsequently calculated based on the imaginary part of Green's function \cite{novotny_principles_2006}.
}

\showmatmethods{} % Display the Materials and Methods section

\acknow{We thank Alessandro Parisotto for providing the specimen and for proof-reading the manuscript. This study was supported by the
European Research Council (ERC) through grant PrISMoID (833895). The authors further acknowledge financial support by the Adolphe Merkle Foundation and the Swiss National Science Foundation (SNSF) through the National Center of Competence in Research \textit{Bio-Inspired Materials}.}

\showacknow{} % Display the acknowledgments section

% Bibliography
\bibliography{biblio}

\newpage
\onecolumn
\pagenumbering{arabic} 
\setcounter{page}{1}
\appendix
\renewcommand{\thefigure}{S\arabic{figure}}
\setcounter{figure}{0}

\section*{Supplementary Information for}
\LARGE{\textbf{3D tomographic analysis of the order-disorder interplay in the \textit{Pachyrhynchus \\ congestus mirabilis} weevil \vspace{3mm}}} \\
\noindent \large{\noindent Kenza Djeghdi, Ullrich Steiner and Bodo D. Wilts* \\}
\normalsize
*Corresponding author, Email: bodo.wilts@plus.ac.at\\

\subsection*{Imaging infiltrated scales} 
\begin{figure}[h!]
\centering
\includegraphics[width=0.9\textwidth]{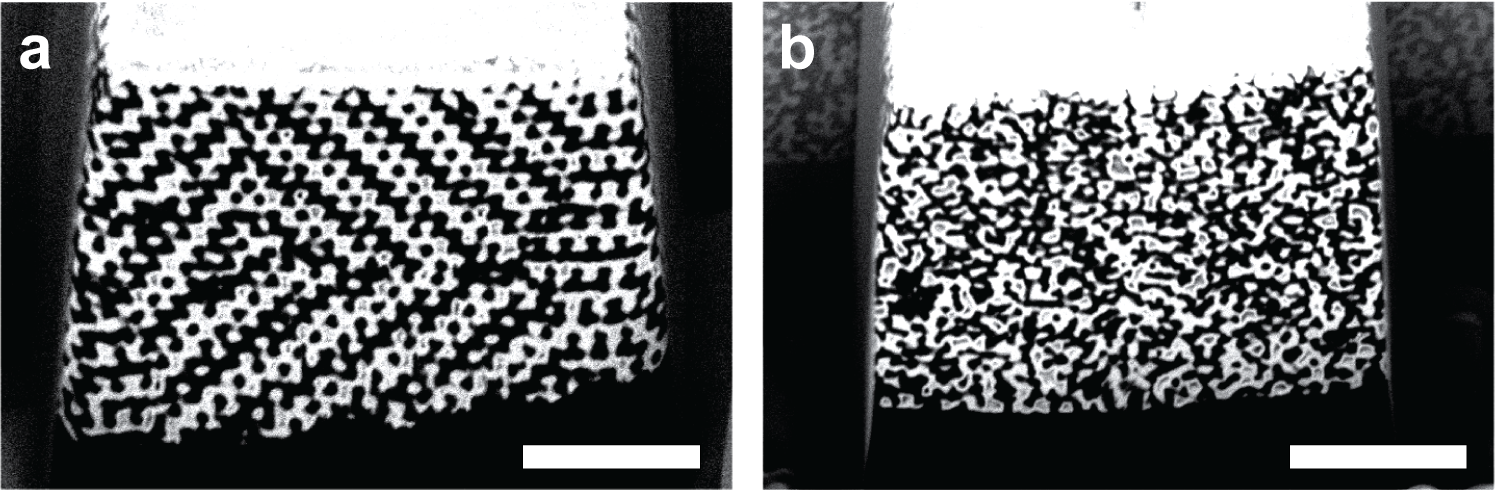}
\caption{SEM image post Pt infiltration of a) an ordered red scale and b) a disordered blue scale. The chitin phase appears in black and the more conductive Pt phase appears in white. Scale bars: 2\,\textmu m.}
\label{fig:pt-ebid}

\end{figure}

Cross-sections of scales (Fig.~\ref{fig:pt-ebid}) were taken by FIB-SEM after in-situ Pt filling and showed a strong contrast between the two phases of the bicontinuous networks when using the T1 detector.

\subsection*{Angle definitions} 

A bond angle is the angle formed between three atoms across two bonds. It is defined by
 
\begin{equation*}
    \theta = \cos^{-1}\left(\frac{\Vec{r_1} \cdot \Vec{r_2}}{\| \Vec{r_1}\| \cdot \| \Vec{r_2}\|}\right). 
\end{equation*}
    
 \vspace{5mm} A (proper) dihedral angle (also called torsion angle) is defined as the angle between two planes which are  delimited by two  bonds emerging from neighboring atoms,    
  \begin{equation*}
    \phi = \text{atan2}(\Vec{r_2} \cdot ((\Vec{r_1} \times \Vec{r_2}) \times (\Vec{r_2} \times \Vec{r_3})), \| \Vec{r_2}\| (\Vec{r_1} \times \Vec{r_2}) \cdot (\Vec{r_2} \times \Vec{r_3})).
\end{equation*} 

\begin{figure}[h!]
\centering
\includegraphics[width=0.5\textwidth]{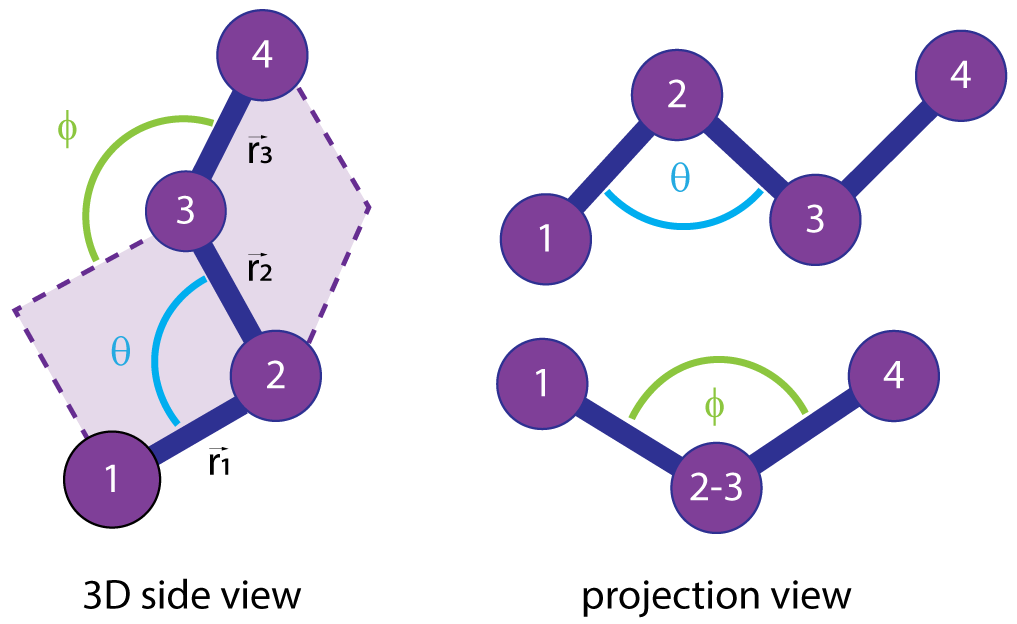}
 \caption{Definition of bond angle $\theta$ (blue) and dihedral angle $\phi$ (green).}
\label{fig:angle_def}
\end{figure}

Representation of both angles is shown Fig.~\ref{fig:angle_def}.

\newpage

\subsection*{Transmission spectra}
\begin{figure}[h!]
\centering
\includegraphics[width=\textwidth]{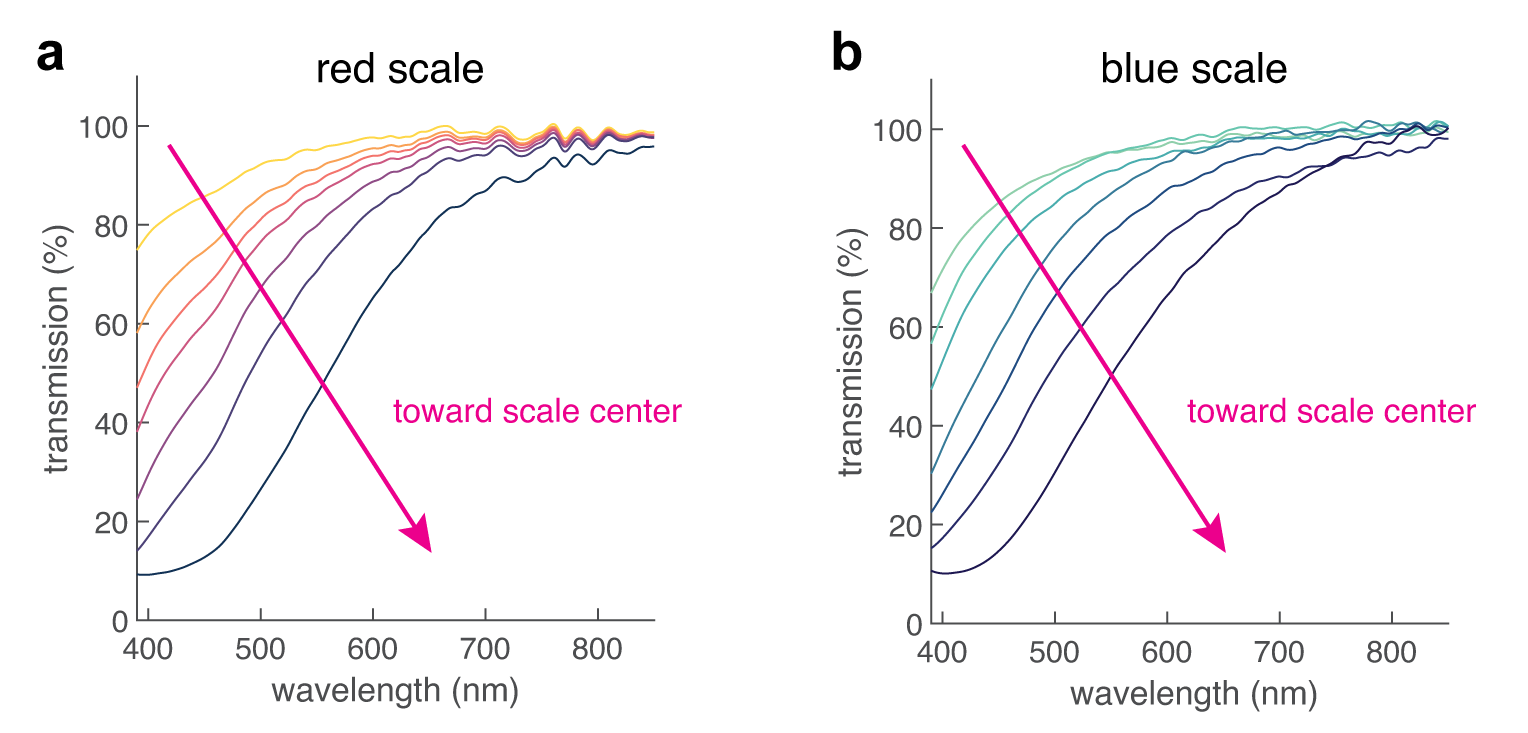}
\caption{Transmission spectra taken at various positions of a red (a) and a blue (b) scale immersed in  a fluid matching the refractive index of chitin. The concentration of pigment increases from the edge of the scale to the stem, giving rise to an absorption gradient.
%The closer to the stem the broader the absorption dip which correlates with the pigment concentration gradient from the center of the scale to the edges.
}
\label{fig:Transmission}
\end{figure}

Transmission spectra (Fig.~\ref{fig:Transmission}) were taken in refractive index matching fluid to eliminate the contribution of structural  coloration, thereby investigating the presence of pigments. The refractive index  was calculated from the Cauchy law\footnote[2]{DG Stavenga, HL Leertouwer, T Hariyama, HA De Raedt, BD Wilts, Sexual Dichromatism of the Damselfly Calopteryx
japonica Caused by a Melanin-Chitin Multilayer in the Male Wing Veins. PLoS ONE 7 (2012).} %\cite{stavenga_sexual_2012}
and the absorption coefficient was calculated from the transmission spectra using
\begin{equation}
    \text{Re}(n_\text{melC})= 1.517 + \frac{8800}{\lambda^2},
    \label{eq:Re(n)}
\end{equation}
\begin{equation}
    \text{Im}(n_\text{melC})= -\log_{10}(\text{T})~\frac{\lambda}{4 \pi \rm d \log_{10} ({\rm e})},
    \label{eq:Im(n)}
\end{equation}
where $T$ is the transmission value, $\lambda$ is the wavelength, and $d$ is the thickness of the scale.

\newpage
\subsection*{3D representation}
\begin{figure}[h!]
\centering
\includegraphics[width=0.85\textwidth]{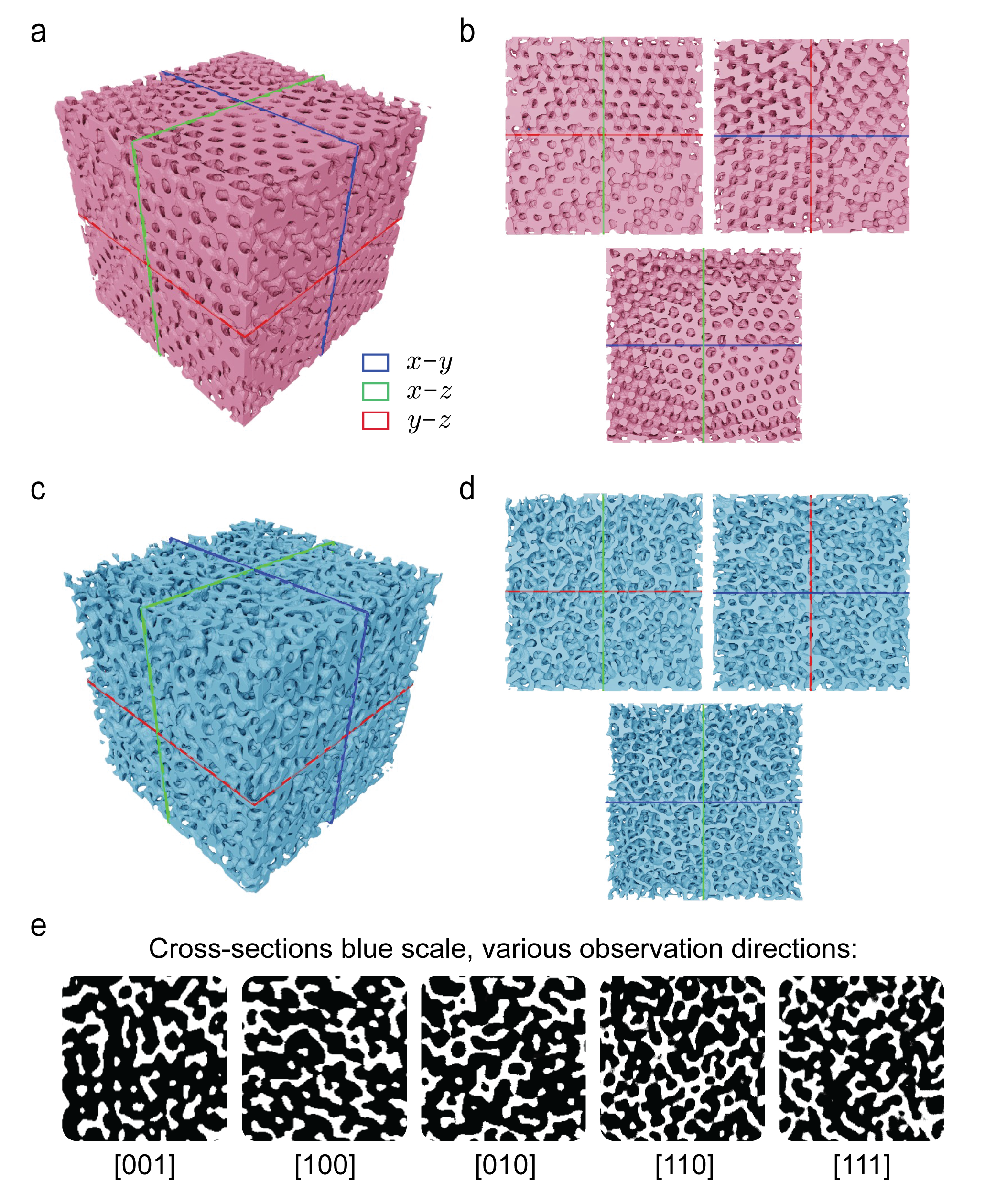}
\caption{Sub-volume $V = (3.6\,\text{\textmu m})^3$ of the a) red and c) blue scales. $x-y$, $x-z$, and $y-z$ planes of the b) red and d) blue scales. e) Cross-sections through the 3D dataset of the blue scale. The Miller indices correspond to the direction of observation in the referential defined above.}
\label{fig:3D}
\end{figure}

Additional representations of the reconstructed sub-volumes are shown in Fig.~\ref{fig:3D}. In this representation, the plane imaged by the SEM is $x-y$, the ion beam is aligned with the $O-x$ axis and the slicing direction is along the $O-z$ axis.

\end{document}